\documentclass[apl, aps, two column,showpacs,pra,floatfix]{revtex4-1}
\usepackage{comment, xcolor, multirow, float,graphicx,epsfig,amssymb,amsmath,epstopdf,amsmath, bbold, mathtools, hyperref, url, physics}
\usepackage[caption=false]{subfig}
\hypersetup{colorlinks=true, citecolor = blue, linkcolor=blue, filecolor=magenta, urlcolor=blue}

\begin{document}
\title{Enhanced quantum illumination of a lossy target: A sequential interaction model}
\author{Shilpi Srivastava}
\email{shilpi.22phz0008@iitrpr.ac.in}
\author{Shubhrangshu Dasgupta}
\affiliation{Department of Physics, Indian Institute of Technology Ropar, Rupnagar, Punjab, 140001, India}

\begin{abstract}
Quantum illumination (QI) exploits quantum correlations to detect weakly reflecting targets embedded in noisy environments with higher sensitivity than classical illumination (CI). We investigate the effectiveness of QI in a realistic setting in which the signal sequentially interacts with a noisy environment and a lossy target. The target is considered at a temperature distinct from its surroundings, while both the interactions are modeled as an action of independent beam splitters with different reflectivities. The detection performance is quantified using the signal-to-noise ratio (SNR) and the quantum Chernoff bound (QCB), the latter providing an upper bound on the error probability. The performance of the Gaussian two-mode squeezed state (TMSS) is compared with that of the optimal classical protocol based on coherent states (CS). The proposed model shows that TMSS consistently achieves a higher SNR than CS for a low-reflectivity target and an arbitrary phase change and remains robust against thermal noise. 
Furthermore, a sufficiently lower QCB is obtained for TMSS than in previously reported results, indicating greater distinguishability between the presence and absence of the target. These findings underscore the role of realistic modeling in improving QI-based detection of lossy targets, with potential relevance to quantum radar and lidar systems.
\end{abstract}

\maketitle
\section{INTRODUCTION}

Illumination of an object using a classical beam of light relies on optical phenomena, namely, scattering, reflection, and refraction. Lloyd has coined the term quantum illumination (QI) \cite{lloyd2008enhanced}, which refers to illumination using quantum properties of light. It has emerged as a paradigmatic shift in visual sensing, which harnesses entanglement to detect the presence of an object immersed in a noisy thermal environment. The basic idea involves preparing a pair of entangled optical modes and sending one of them, the signal, to irradiate the target while storing the other mode, the idler, locally. Upon the return of the signal, a joint measurement on the signal and idler beams is performed to infer the presence or absence of the desired target. In contrast, in the so-called classical illumination (CI) protocol, one does not use any idler mode, and only the signal mode is employed. In QI, the quantum correlations between the two modes enable more reliable discrimination of the presence of the target, either through improved detection accuracy or reduced resource requirements~\cite{shapiro2009quantum,sacchi2005entanglement, tan2008quantum,shafi2023quantum}. This technique has also been implemented for optical-frequency probes, which are relevant for quantum lidar applications~\cite{murchie2021theoretical,murchie2024quantum,li2021dynamic,slaiman2026design,gregory2020imaging}, as well as for microwave-frequency probes used in quantum radar~\cite{barzanjeh2015microwave, barzanjeh2020microwave,bourassa2020progress, luong2020entanglement,   karsa2024quantum,lanzagorta2012quantum,shapiro2020quantum,slepyan2021quantum,torrome2020introduction,torrome2024advances,russer2021performance}. These technologies have been proven to detect weakly reflecting or cloaked targets~\cite{las2017quantum,davuluri2025optical} that are otherwise difficult to detect using the CI schemes.

In QI, the Gaussian two-mode state (TMSS)~\cite{tan2008quantum} has emerged as a prominent candidate, whose performance is typically compared with that of the optimal CI protocol based on the coherent state (CS). The QI protocol has also been experimentally demonstrated using TMSS, establishing it as a practical and experimentally viable resource for target detection \cite{guha2009gaussian,lopaeva2013experimental,zhuang2017optimum}. Notably, the associated advantage persists even when the entanglement between the signal and idler modes is completely degraded by environmental noise~\cite{sacchi2005entanglement, zhang2013entanglement, zhang2015entanglement}.
Although entanglement is widely regarded as the primary resource responsible for quantum advantage, it has been shown that it is not the sole contributor to the robustness of quantum illumination~\cite{kim2023entanglement}. In particular, the persistence of the advantage even in regimes where the entanglement is destroyed has been attributed to other forms of quantum correlations, such as quantum discord~\cite{weedbrook2016discord,bradshaw2017overarching}. Beyond the TMSS, various non-classical states have been investigated, including asymmetrically squeezed two-mode states, high-dimensional Bell states, and non-Gaussian states obtained via photon addition and subtraction of the TMSS~\cite{jo2021quantum,pannu2024quantum,fan2018quantum,gupta2024quantum}. In certain regimes, these states are found to yield lower error probabilities than the Gaussian TMSS. However, the TMSS remains the optimal resource when constraints are imposed on the signal energy. These states have further been analyzed under various types of noise and experimental imperfections~\cite{gupta2024quantum}, providing deeper insight into their practical performance.  

\begin{figure*}
    \subfloat[\label{fig:1a}]{
        \includegraphics[width=0.48\linewidth]{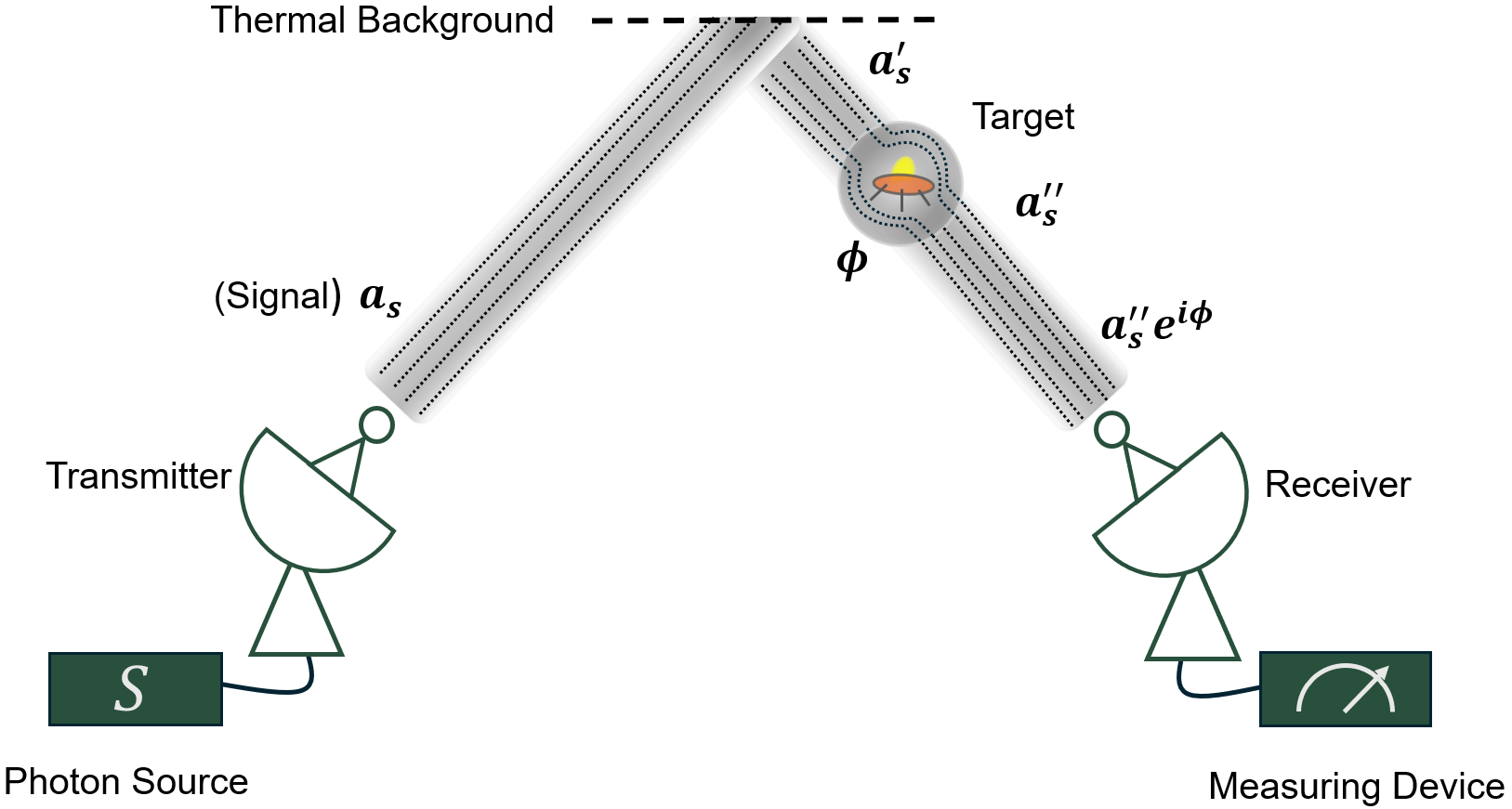}
    }
    \hfill
    \subfloat[\label{fig:1b}]{
        \includegraphics[width=0.48\linewidth]{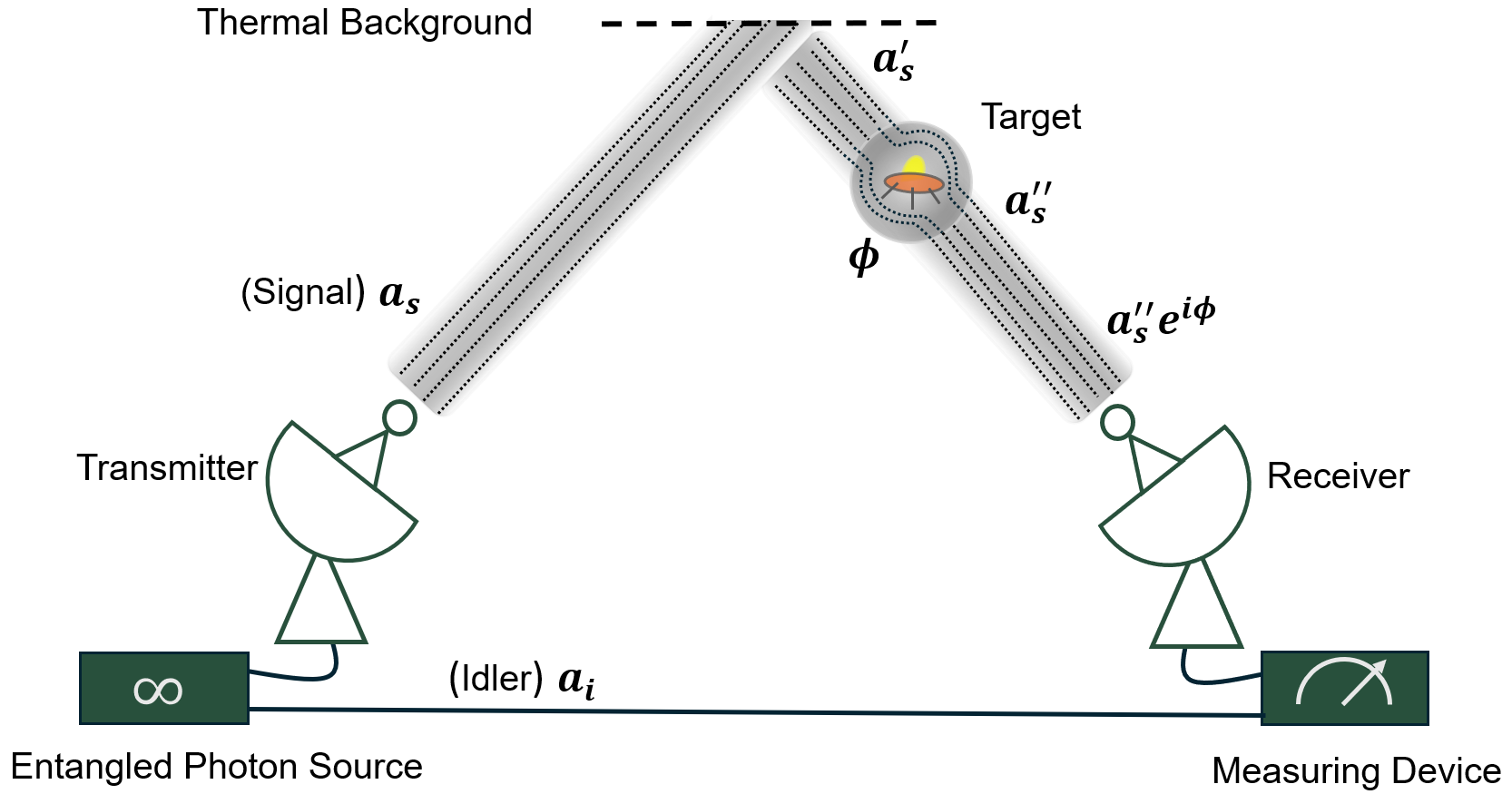}
    }
    \caption{Schematic diagram of (a) CI and (b) QI of a lossy target}
    \label{fig:1}
\end{figure*}

The above studies rely on simplified models of interaction with the target and the environment, underscoring the importance of developing more refined models for practical quantum sensing applications. Recent efforts have focused on incorporating more realistic features into the QI framework. For instance, the effect of target absorption has been modeled using lossy beam splitters~\cite{gupta2021quantum}, revealing that absorption can, perhaps unexpectedly, enhance detection efficiency. Alternative detection scenarios have also been explored, such as phase-shift-based target identification in cloaked environments, where quantum correlations can lead to improved signal-to-noise ratios (SNRs) compared to classical approaches~\cite{las2017quantum}. In addition, the effectiveness of QI is typically quantified using several key metrics, including the signal-to-noise ratio (SNR)~\cite{lopaeva2013experimental, las2017quantum}, the minimum error probability given by the Helstrom limit~\cite{helstrom1969quantum, nair2020fundamental} and the quantum Chernoff bound (QCB)~\cite{calsamiglia2008quantum,nussbaum2009chernoff,pirandola2008computable,kargin2005chernoff,audenaert2007discriminating}, and quantum estimation approaches based on the quantum Fisher information (QFI)~\cite{petz2011introduction,sanz2017quantum}. 

Despite these advances, most existing models treat the interaction of the signal with the environment and the target in an effectively combined manner. In particular, some approaches consider only the mixing of thermal noise with the signal while modeling the target as perfectly reflective~\cite{las2017quantum}. Others incorporate target absorption by modeling it as a beam splitter (BS), but include environmental noise only through one input port, effectively neglecting the intrinsic thermal contribution of the target by assuming it to be in equilibrium with the surrounding environment~\cite{gupta2021quantum, gupta2024quantum}. Such assumptions may not adequately capture realistic physical conditions.

In a more realistic scenario, the signal's interactions with the environment and with the target should be treated as distinct processes, since the target may neither be perfectly reflective nor necessarily at the same temperature as the surrounding environment. Motivated by this observation, we propose a sequential interaction model in which the signal first interacts with the environment and subsequently with the target, each interaction described by an independent beam splitter with a different reflectivity. Furthermore, the thermal modes associated with the environment and the target are treated independently, each characterized by its own average thermal photon number. Within this generalized framework, the performance of CI employing the CS and QI employing the TMSS is investigated. The detection performance is analyzed using the SNR, while the QCB is employed to quantify the error probability associated with target detection. The main contributions of this work are therefore threefold: first, a realistic sequential interaction model is developed in which the environment and the target are treated as distinct physical systems; second, the effect of an independent thermal target is explicitly incorporated by assigning separate thermal modes to the environment and the target; and third, the performance of the proposed model is systematically evaluated through both the SNR and the QCB.

This paper is organized in the following manner. In Sec.~\ref{II.A} and Sec.~\ref{II.B} we introduce the proposed model and the corresponding mathematical framework for evaluating the SNR and the QCB for both classical and quantum protocols. In Sec.~\ref{III} and Sec.~\ref{IV}, we present a comparative analysis of the considered states using these two performance metrics. We conclude the paper in Sec. \ref{V}.

\section{Model and performance measures}
We analyze and compare the efficiency of CI and QI schemes for a lossy or partially reflecting target based on the performance metrics SNR and QCB, following the methodology outlined in \cite{las2017quantum,fan2018quantum}. The schematic representations and the equivalent circuit of both the protocols are shown in Figs.~\ref{fig:1} and~\ref{fig:2}, respectively. The key difference between the CI and QI protocols lies in the use of an idler mode in QI, which is prepared in an entangled state with the signal mode and subsequently used for joint detection.

\subsection{Sequential interaction model}
\label{II.A}

\begin{figure*}
    \subfloat[\label{fig:2a}]{
        \includegraphics[width=0.48\linewidth]{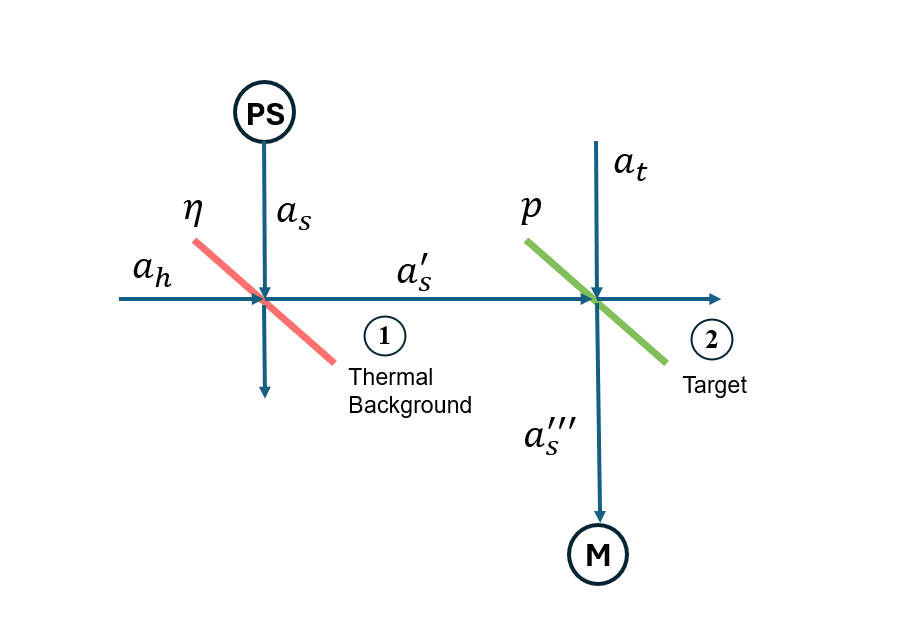}
    }
    \hfill
    \subfloat[\label{fig:2b}]{
        \includegraphics[width=0.48\linewidth]{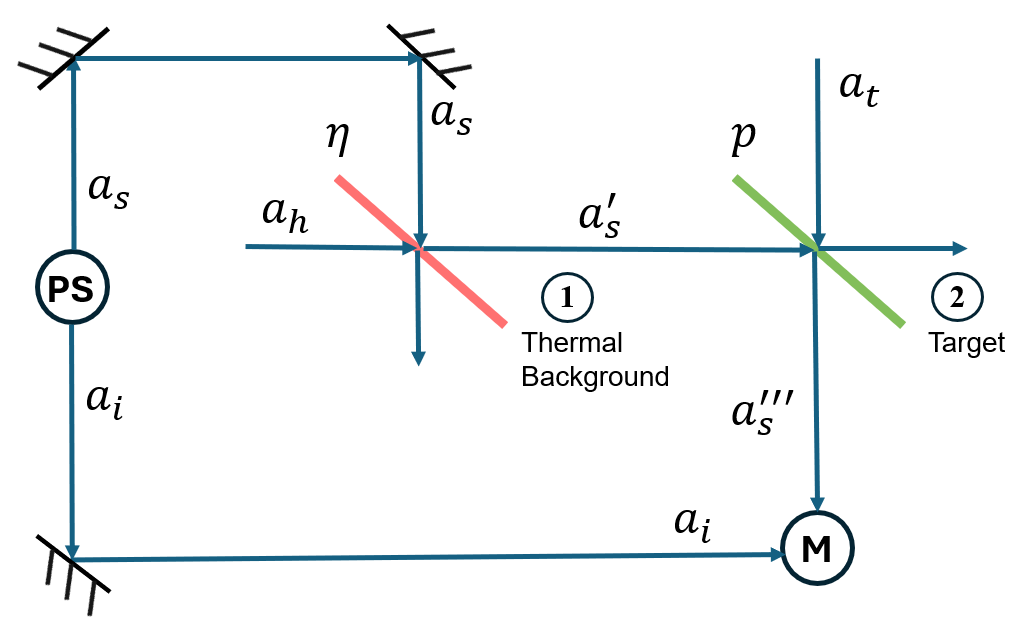}
    }
    \caption{Equivalent circuit of (a) CI and (b) QI of a lossy target. Here, PS: Photon Source and M: Measurement device}
    \label{fig:2}
\end{figure*}

In the case of CI, as shown in Fig.\ref{fig:1a} and Fig.~\ref{fig:2a}, the signal beam is prepared in the laboratory and transmitted through the transmitter to search for the target. The signal can be expressed in terms of the bosonic annihilation operator of the photon field, $a_s$. Initially, the signal emitted from the transmitter interacts with the bright thermal environment. This interaction is modeled as the action of a high-reflectivity BS (marked as 1 in Fig.~\ref{fig:2a}) with reflectivity $\eta$ and transmissivity (\(1- \eta \)). As a result, the signal mode is transformed according to the BS interaction as
\begin{equation}
    a^{\prime}_s = \sqrt{\eta} \, a_s + \sqrt{1 - \eta} \, a_h, \label{1}
\end{equation}  
where \( a^{\prime}_s \) denotes the modified annihilation operator of the transmitted signal and \( a_h \) represents the annihilation operator of the environmental (thermal) noise. 

Next, when the lossy target is present, a portion of the intermediate signal \( a^{\prime}_s \) is reflected after mixing with the noise from the target. In real-world scenarios where quantum radars and lidars are deployed, the target typically absorbs or scatters most of the signal in arbitrary directions, reflecting only a small fraction. Therefore, we model the interaction with the target as an action of another BS (marked as 2 in Fig.~\ref{fig:2a}) with small reflectivity \( p \) and transmissivity (\(1-p \)), resulting in the transformation  
\begin{equation}
    a^{\prime\prime}_s = \sqrt{p} \, a^{\prime}_s + \sqrt{1 - p} \, a_t,
\end{equation}
where \( a_t \) denotes the annihilation operator of the target's intrinsic thermal mode. Note that for a fully reflecting target, $p=1$. Therefore, the model presented here is a natural extension of the model of ~\cite{las2017quantum}, with their result obtained as the special case $p=1$. 

Finally, the signal \( a^{\prime\prime}_s \) acquires a phase shift \( \phi \) after being deflected by the target,  
\begin{equation}
    a^{\prime\prime\prime}_s = a^{\prime\prime}_s e^{i\phi}.
\end{equation}

Expressed in terms of \( a_s \), \(a_h \), and \(a_t \), the final form of the signal becomes  
\begin{equation}
    a^{\prime\prime\prime}_s = \left[\sqrt{p\eta} \, a_s + \sqrt{p(1 - \eta)} \, a_h + \sqrt{1 - p} \, a_t \right] e^{i\phi}.
    \label{4}
\end{equation}


Now, in the QI protocol, as shown in Fig.~\ref{fig:1a} and Fig.~\ref{fig:2b}, an entangled signal-idler photon pair is generated. The annihilation operator of the generated signal is again represented by $a_s$, whereas $a_i$ denotes the idler mode. The signal photon propagates through the thermal background, interacts with the target, and undergoes the same sequence of interactions as in the CI protocol. The modified signal then returns to the receiver, where it is this time jointly measured with the idler photon retained throughout the above process.

The corresponding two-mode mixing operators describing the BS operation for the BS 1 and BS 2 are $U_1(\zeta) = \exp(\zeta {a_s}^\dagger a_h - \zeta^* a_s {a_h}^\dagger)$ and $U_2({\zeta}^\prime) = \exp({\zeta}^\prime {a_s^\prime}^\dagger a_t - {\zeta^\prime}^* a_s^\prime {a_t}^\dagger)$ respectively, where $\zeta=\arcsin\sqrt{\eta}$ and $\zeta^\prime = \arcsin\sqrt{p}$. The $a_s'$ denotes the signal mode after the interaction with the first BS and is defined by Eq.~(\ref{1}). 

\subsection{The SNR and the QCB}
\label{II.B}

The SNR is a measure to quantify how well a signal can be distinguished from background noise. It is commonly defined as the ratio of the square of the mean value of the measured signal to its variance, thereby capturing the relative strength of the signal against fluctuations~\cite{loudon1974quantum}. A higher SNR indicates that the signal is more clearly discernible, whereas a lower SNR implies that it is significantly affected by noise. As described in Sec. ~\ref{III}, we calculate the SNR by measuring the quadratures of the signal $a^{\prime\prime\prime}_s$ alone in the CI protocol. On the other hand, in QI, we need to perform a joint quadrature measurement of the signal and idler beams to calculate the SNR.
In the context of QI, the SNR reflects the strength of correlations between the received signal and the retained idler in the presence of environmental noise. It provides a useful metric for comparing the performance of classical and quantum strategies, particularly in situations where the signal is weak and the background noise is substantial.

To calculate the SNR in the QI, one needs to perform a joint continuous-variable Bell measurement. For such measurement, one typically first makes local homodyne measurements of rotated quadratures, $X_j(\theta_j)=x_j\cos\theta_j + p_j\sin\theta_j$ ($j\in s,i$). For $\theta_j = 0$ and $\frac{\pi}{2}$, this quadrature will correspond to the position quadrature $x_j$ and the linear momentum quadrature $p_j$, respectively. Therefore, for bipartite correlations, appropriate choices of the local phases $\theta_j$ allow one to access combinations of moments such as $x_s x_i \pm p_s p_i$ and $x_s p_i \pm p_s x_i$~\cite{cavalcanti2007bell,he2010bell}. The SNR is then defined as the ratio of the expectation values of these combinations and their variance~\cite{las2017quantum}. For the QI, one makes such measurements for a signal beam $a_s^{\prime\prime\prime}$ received and the idler beam $a_i$ retained at the laboratory. In the case of CI, since there is no idler beam, one calculates the SNR of the signal $a_s^{\prime\prime\prime}$ alone, without requiring quadrature rotation. 

Another key performance measure is the QCB, which provides a computable upper bound on the minimum error probability for discriminating between the target’s presence and absence in the presence of unavoidable thermal noise. It determines how rapidly the error probability decreases as more measurements are performed. A smaller value of the QCB implies a faster decrease of the error probability, and therefore better discrimination between the two hypotheses. When no target is present (hypothesis: $H_0$), the transmitted signal is completely lost, and the detector only records background contributions. In contrast, if the target is present (hypothesis: $H_1$), a fraction of the transmitted signal is reflected and received together with the thermal noise. 
The problem of determining whether the target is present or absent reduces to a binary quantum state discrimination task \cite{barnett2009quantum, calsamiglia2010local}. The performance of this discrimination is quantified by the minimum error probability associated with distinguishing between the two states, \( \rho_0 \) and \( \rho_1 \) corresponding to hypotheses $H_0$ and $H_1$ respectively, with a certain a priori probability. 
Depending on the absence or presence of the target, the measurement device M receives one of the two possible output states, denoted by \( \rho_0 \) and \( \rho_1 \), respectively. In our model, we define \( \rho_0 \) and \( \rho_1 \) as:
\begin{align} 
&\rho_0 = \mathrm{Tr}_{st}\!\Big[
U_2(p=0)\,
\Big(
U_1(\rho_{\rm in}\!\otimes\!\rho_h)U_1^\dagger
\otimes \rho_t
\Big) \nonumber \\
&\qquad\qquad
U_2^\dagger(p=0)
\Big], \\[6pt]
&\rho_1 = \mathrm{Tr}_{ht}\!\Big[
U_2(p>0)\,
\Big(
U_1(\rho_{\rm in}\!\otimes\!\rho_h)U_1^\dagger
\otimes \rho_t
\Big) \nonumber \\
&\qquad\qquad
U_2^\dagger(p>0)
\Big].
\end{align}

Here, $U_2(p=0)$ reduces to the identity operator, indicating the absence of any interaction with the target. For QI, the probe state is an entangled signal–idler state $\rho_{\text{in}}=\rho_{is}$, whereas for CI, the probe reduces to a single-mode signal state $\rho_{\text{in}}=\rho_s$. The density operators $\rho_h$ and $\rho_t$ describe the thermal states of the environment and the target background with mean photon numbers $n_{th}$ and $n_t$, respectively.

For the hypothesis \(H_0\), only the signal-environment interaction \(U_1\) is applied. 
The target mode is included as an auxiliary mode for notational consistency, but it is traced out and does not contribute to the received state. Thus, under \(H_0\), the receiver collects the environmental return mode. For the hypothesis \(H_1\), the signal first interacts with the environment through \(U_1\) and subsequently with the target through \(U_2\). After tracing out the unused environmental and target output modes, the receiver collects the returned signal mode.
In the QI protocol, the returned mode is measured jointly with the retained idler, whereas in the CI protocol, only the return mode is measured.

Under the assumption that the target is equally probable to be present or absent, the minimum error probability \( P_e \) for \( M \) identical copies of the entangled probe states is given by the Helstrom limit \cite{helstrom1969quantum}:
\begin{equation}
    P_{e, M} = \frac{1}{2}\left( 1 - \frac{1}{2} \lVert \rho^{\otimes M}_0 - \rho^{\otimes M}_1 \rVert \right).
\end{equation}
Since evaluating \( P_e \) from the above expression is analytically intractable, we instead focus on its asymptotically tight upper bound, which is easier to evaluate and is given by
\begin{equation}
    P_{e,M} \leqslant \xi^M_{QCB}=\frac{1}{2}\left\{\min _{0 \leqslant b \leqslant 1} \operatorname{Tr}\left[\rho_0^{\mathrm{\textit{b}}} \rho_1^{1-\mathrm{\textit{b}}}\right]\right\}^M,
\end{equation}
where the quantity \(\xi^M_{QCB}\) is called the QCB \cite{audenaert2007discriminating}. For a single copy of the probe state,
\begin{equation}
    \xi_{QCB}=\frac{1}{2}\left\{\min _{0 \leqslant b \leqslant 1} \operatorname{Tr}\left[\rho_0^{\mathrm{\textit{b}}} \rho_1^{1-\mathrm{\textit{b}}}\right]\right\}.
    \label{9}
\end{equation}

The evaluation of the QCB depends strongly on the structure of the quantum states involved. A variety of analytical and numerical approaches have been developed in the literature. For Gaussian states, which are commonly encountered in the QI protocols, the QCB can be computed analytically in terms of covariance matrices and displacement vectors, leading to closed-form expressions that significantly simplify the analysis \cite{tan2008quantum,pirandola2008computable,gupta2024quantum}. In contrast, for more general non-Gaussian or mixed states, the direct evaluation of the QCB typically requires numerical optimization over the Chernoff parameter, as the calculation of fractional powers of density operators becomes analytically intractable~\cite{fan2018quantum}.

In the present work, we evaluate the QCB numerically by performing the minimization over the Chernoff parameter. This approach provides a reliable estimate of the error exponent while capturing the full complexity of the model.

\section{COMPARATIVE STUDY FOR THE SNR} 

\begin{figure*}
\centering

\subfloat[]{%
    \includegraphics[width=0.32\textwidth]{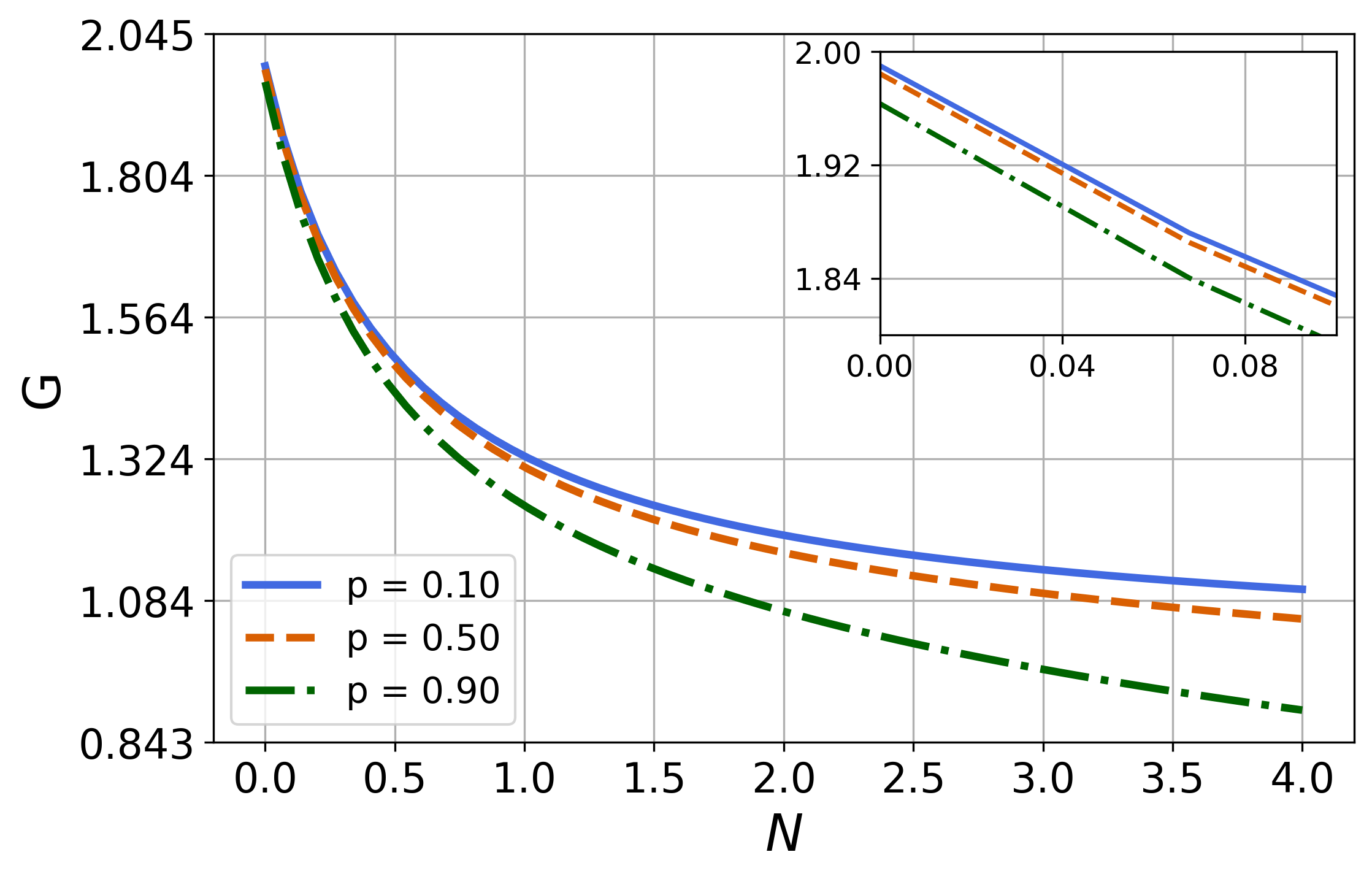}
    \label{fig:3a}
}
\hfill
\subfloat[]{%
    \includegraphics[width=0.32\textwidth]{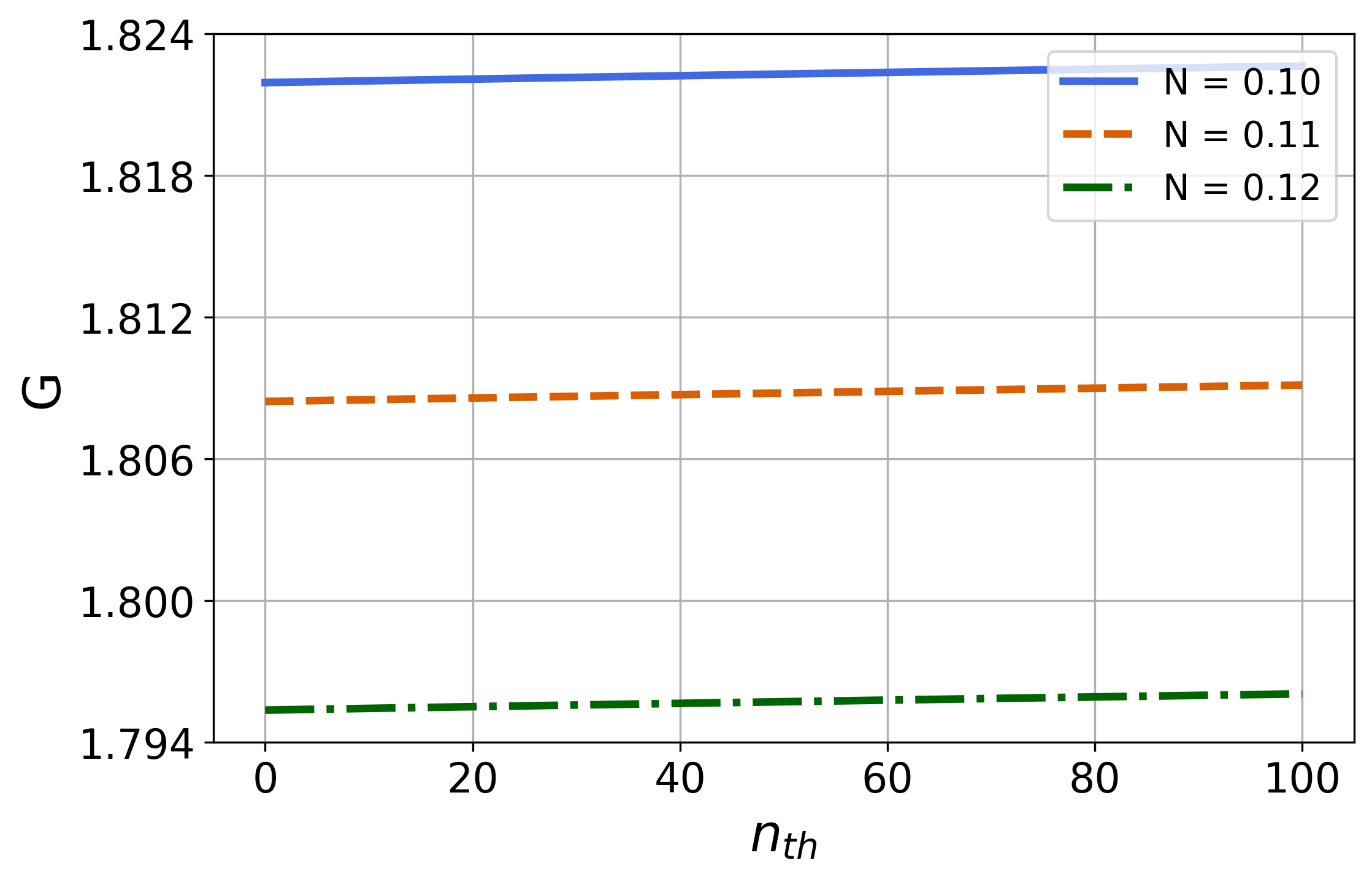}
    \label{fig:3b}
}
\hfill
\subfloat[]{%
    \includegraphics[width=0.32\textwidth]{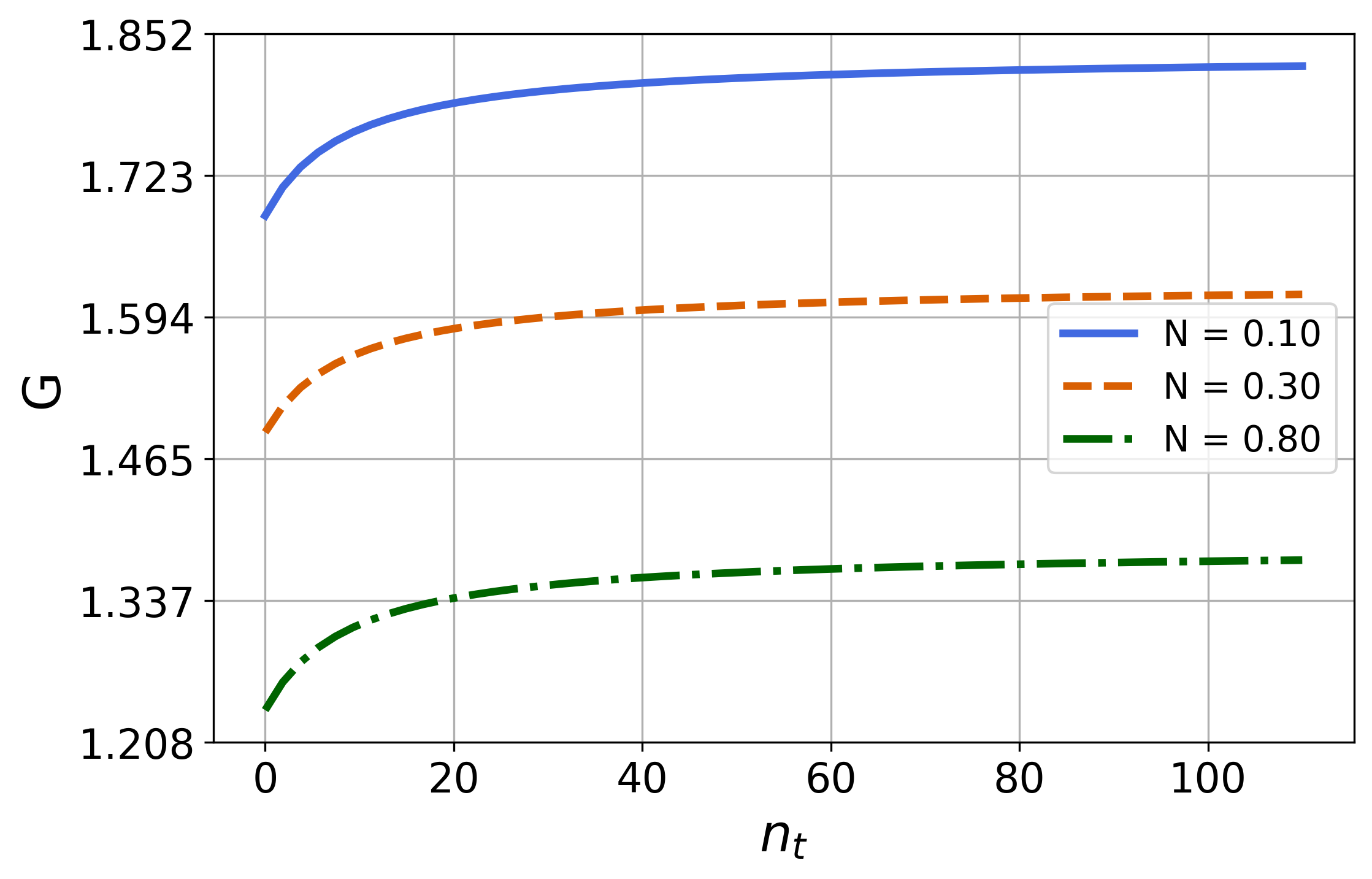}
    \label{fig:3c}
}

\vspace{0.5cm}

\subfloat[]{%
    \includegraphics[width=0.32\textwidth]{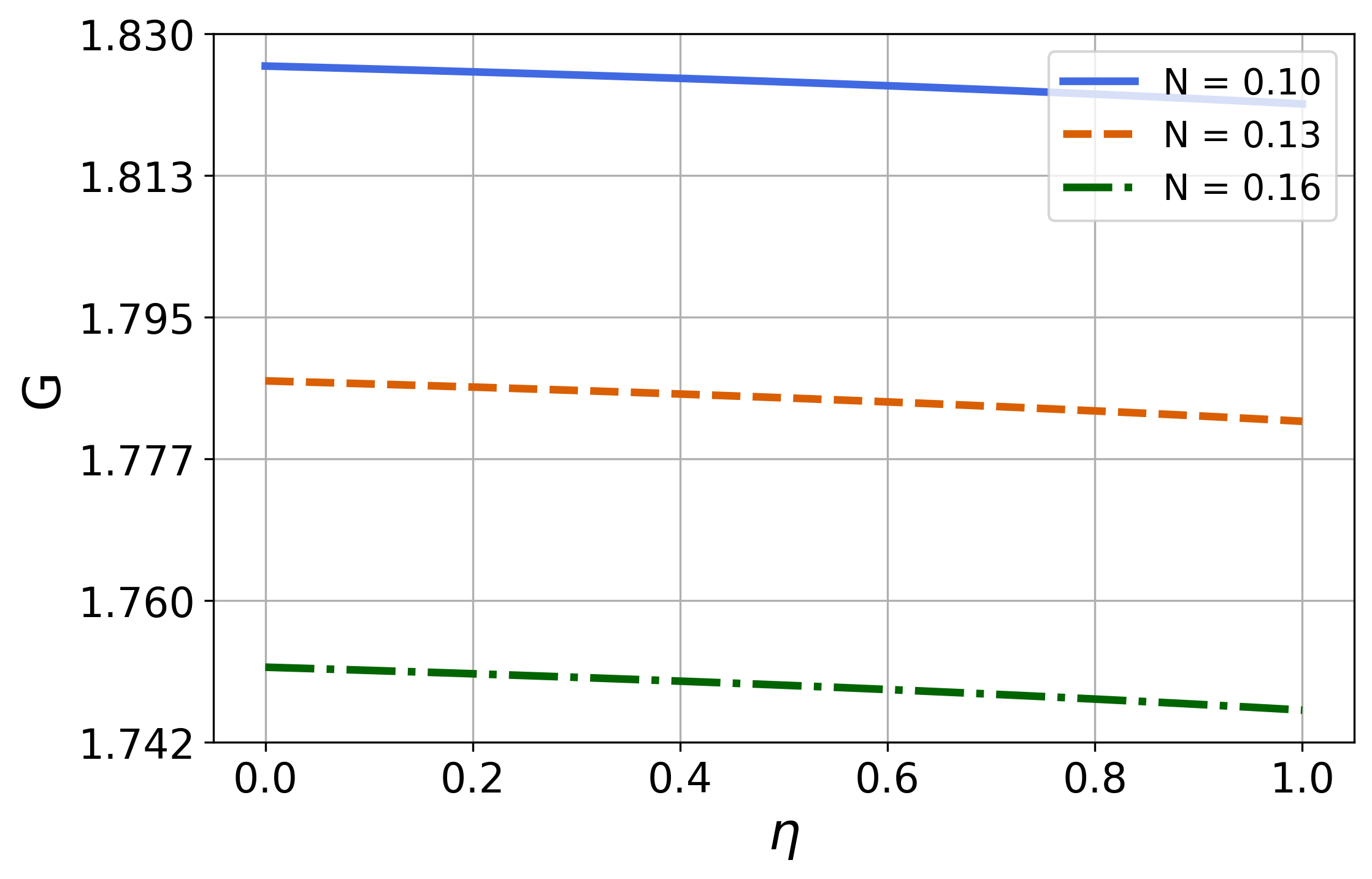}
    \label{fig:3d}
}
\hfill
\subfloat[]{%
    \includegraphics[width=0.32\textwidth]{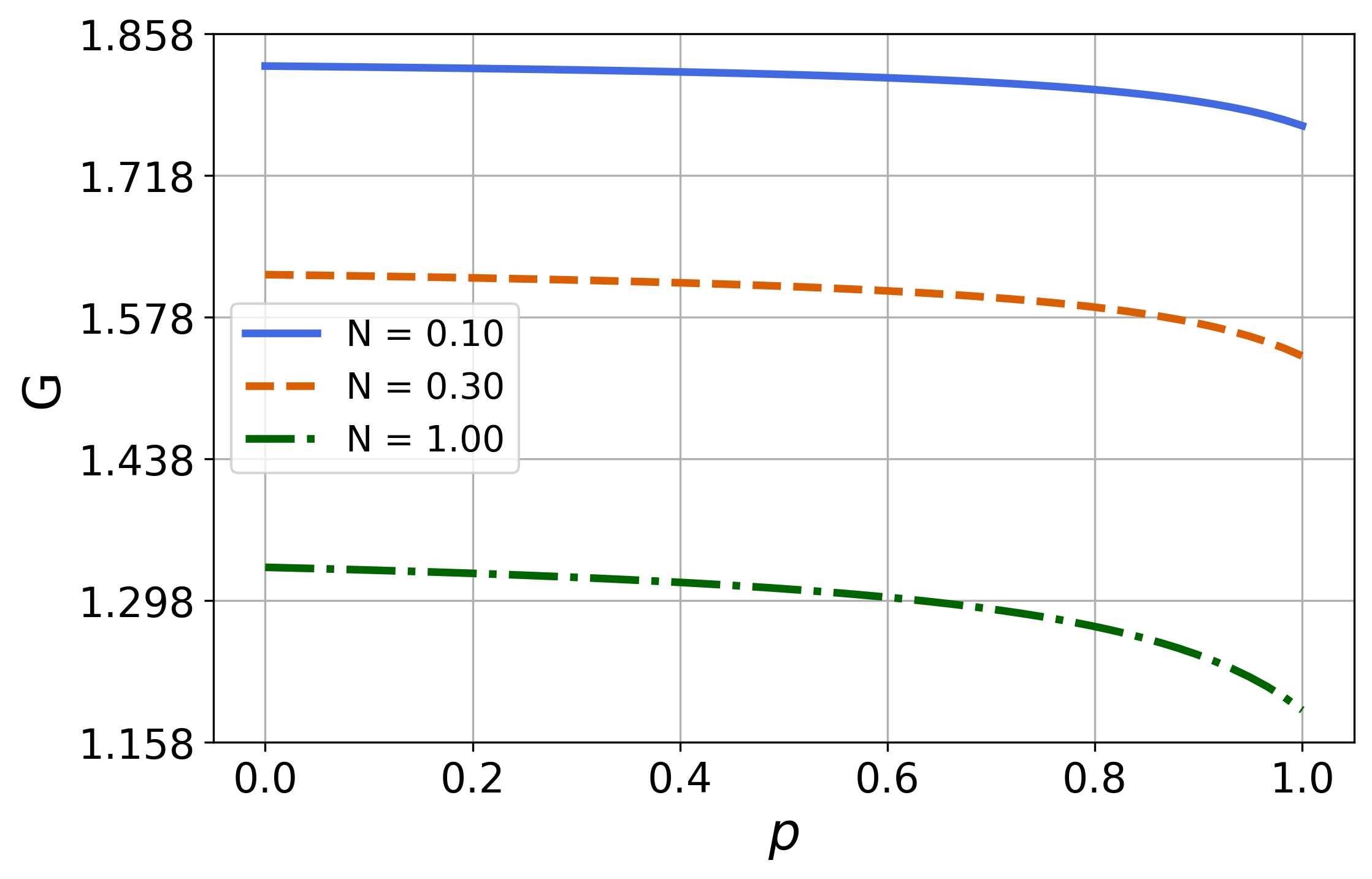}
    \label{fig:3e}
}
\hfill
\subfloat[]{%
    \includegraphics[width=0.32\textwidth]{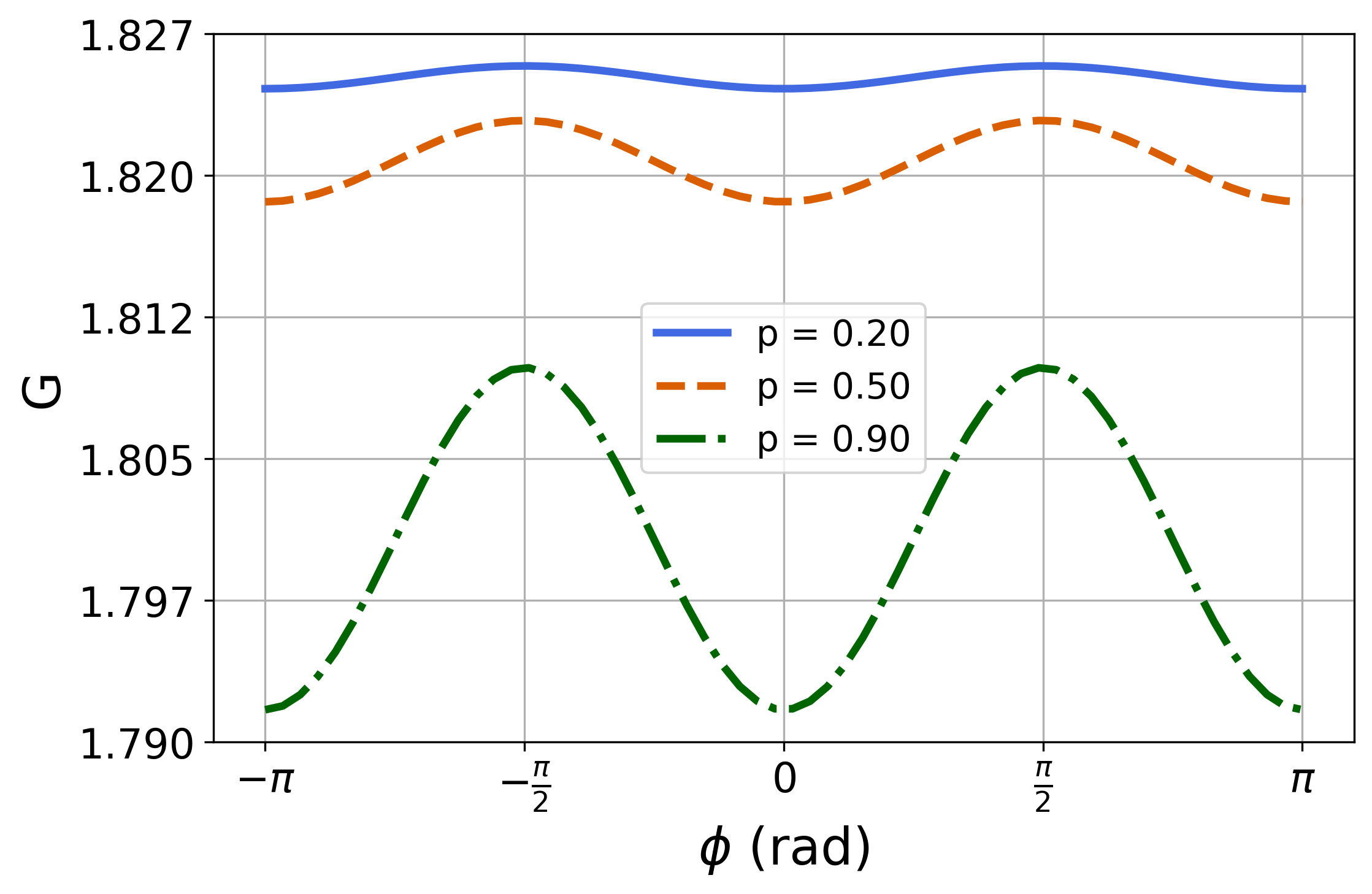}
    \label{fig:3f}
}

\caption{The gain $G$ as a function of (a) $N$, (b) $n_{th}$, (c) $n_t$, (d) $\eta$, (e)  $p$, and (f) $\phi$ for $N=0.1$. Unless otherwise stated, the parameters chosen are $\eta = 0.8$, $p=0.3$, $n_{th} = 100$, $n_t = 110$, and $\cos^2\phi = 1$.  }
\label{fig:3}
\end{figure*}

\label{III}
\subsection{Classical Protocol}
\label{III.A}
We begin by considering the optimal classical strategy for detecting small phase shifts. A coherent state is expressed as
\begin{equation}
    |\alpha\rangle = e^{-|\alpha|^2/2} \sum_{n=0}^{\infty} \frac{\alpha^n}{\sqrt{n!}} |n\rangle\;.\label{10}
\end{equation} 
After interacting with the environment and the target, and undergoing a possible phase shift, the appropriate observable to measure is the field quadrature aligned with the coherent amplitude. Assuming that the amplitude $\alpha$ is real without loss of generality, this measurement corresponds to the expectation value of quadrature $x_s^{\prime\prime\prime}$ and the corresponding variance, which is expressed in terms of annihilation and creation operators as
\begin{equation}
  \langle x_s^{\prime\prime\prime} \rangle = \frac{1}{\sqrt{2}} \left\langle a_s^{\prime\prime\prime} + a_s^{\dagger\prime\prime\prime} \right\rangle,
\end{equation}
and
\begin{equation}
   \langle (x_s^{\prime\prime\prime})^2 \rangle - \langle x_s^{\prime\prime\prime} \rangle^2 
   = \frac{1}{2} \left[ \left\langle \left(a_s^{\prime\prime\prime} + a_s^{\dagger\prime\prime\prime}\right)^2 \right\rangle 
   - \left\langle a_s^{\prime\prime\prime} + a_s^{\dagger\prime\prime\prime} \right\rangle^2 \right],
\end{equation}
where $\langle \cdot \rangle=\bra{\alpha} \cdot \ket{\alpha}$ denotes the expectation value taken over the CS.

By expressing $a_s^{\prime\prime\prime}$ in terms of $a_s$, $a_h$, and $a_t$ using Eq.~(\ref{4}), the resulting SNR for the CI is obtained as
\begin{equation}
SNR\,_C = \frac{\langle \bar{x}_s^{\prime\prime\prime} \rangle^2}{\langle (x_s^{\prime\prime\prime})^2 \rangle - \langle x_s^{\prime\prime\prime} \rangle^2} 
= \frac{2 \eta p N (1 - \cos \phi)^2}{p(1-\eta)n_{th} + (1-p)n_t + \frac{1}{2}}\,.
\label{11}
\end{equation}

Here, $\langle \bar{x}_s^{\prime\prime\prime} \rangle = \langle x_s^{\prime\prime\prime} \rangle |_\phi - \langle x_s^{\prime\prime\prime} \rangle $ ensures that the SNR is zero when the de-phased angle $\phi = 0 $. In this expression, $N = |\alpha|^2$ is the average number of photons in the input coherent state, $n_{th}$ and $n_t$ denote the average number of thermal photons in the environment and the target, respectively. We assume that the target is not in thermal equilibrium with the environment; therefore, $n_{t}$ need not equal $n_{th}$.

\subsection{Quantum Protocol} 
\label{III.B}
We consider a TMSS for the quantum protocol. A Gaussian TMSS is defined as  
\begin{equation}
|\Psi\rangle = \sqrt{1 - \lambda^2} \sum_{n=0}^{\infty} \lambda^n |n\rangle_s |n\rangle_i\;,
\label{12}
\end{equation}  
where the subscripts $s$ and $i$ refer to the signal and idler modes, respectively. The parameter \( \lambda \) denotes the squeezing strength, and the average photon number in each mode is given by \( N = N_s = N_i = \frac{\lambda^2}{1 - \lambda^2} \).

In this scheme, the signal mode undergoes the same transformation \( a_s \rightarrow a_s^{\prime\prime\prime}\)  as in the classical protocol, while the idler mode \( a_i \) remains unaltered in the laboratory. 
The SNR in this case is obtained by measuring the combination of the rotated quadratures, $\langle x_s^{\prime\prime\prime} x_i - p_s^{\prime\prime\prime} p_i \rangle$ and its corresponding variance. The quadrature operators for the signal and idler modes are defined in terms of the annihilation and creation operators as
\begin{eqnarray}
&&x_s''' = \frac{1}{\sqrt{2}}(a_s'''^\dagger + a_s'''), \quad 
p_s''' = \frac{i}{\sqrt{2}}( a_s'''^\dagger - a_s'''),
\\
&&x_i = \frac{1}{\sqrt{2}}(a_i^\dagger + a_i), \quad 
p_i = \frac{i}{\sqrt{2}}(a_i^\dagger - a_i),
\end{eqnarray}
respectively. The primed operators indicate the successive transformations undergone by the signal mode during its interaction with the environment and the target. The SNR for the QI protocol is thus defined as
\begin{equation}
\mathrm{SNR}_Q = \frac{\expval{\bar{x}_s''' \bar{x}_i - \bar{p}_s''' \bar{p}_i}^2}
{\expval{(x_s''' x_i - p_s''' p_i)^2} - \expval{x_s''' x_i - p_s''' p_i}^2} \,,
\end{equation}
where the numerator represents the squared mean value of the measured observable, and the denominator corresponds to its variance. Again, $\expval{\bar{x}_s''' \bar{x}_i - \bar{p}_s''' \bar{p}_i} = \left.\expval{x_s''' x_i - p_s''' p_i}\right|_{\phi=0} - \expval{x_s''' x_i - p_s''' p_i}$ ensures that the SNR vanishes when $\phi = 0$.

The average $\langle \cdot \rangle$ in this case is taken over the TMSS, i.e., $\langle \cdot \rangle = \bra{\Psi} \cdot \ket{\Psi}$. The final expression for the SNR in the QI protocol is obtained as follows:  

\begin{widetext}
\begin{equation}
SNR\,_Q = 
\frac{4p \eta N (N + 1) (1 - \cos \phi)^2}{
p \eta \left( 1 + 4N(N+1)\cos^2 \phi \right) + p(1 - \eta)\left( 2n_{th} N + n_{th} + N + 1 \right) + (1 - p)\left( 2n_t N + n_t + N + 1 \right)}\label{13}
\end{equation}
\end{widetext}

To evaluate the advantage of the quantum protocol over the classical one, we consider the ratio of their respective SNRs and define it as the gain $G$, given by
\begin{widetext}
\begin{equation}
G =
\frac{SNR_Q}{SNR_C}
=
\frac{2(N+1)\left(p(1-\eta)n_{th} + (1-p)n_t + \frac{1}{2}\right)}
{p \eta \left(1 + 4N(N+1)\cos^2 \phi \right) + p(1- \eta)\left(2n_{th} N + n_{th} + N+1\right) +(1-p)\left(2n_t N + n_t + N+1\right)}\label{14}
\end{equation}
\end{widetext}

Note that for $p=1$, the target becomes perfectly reflecting and the results in the Eqs.~(\ref{11}),~(\ref{13}), and~(\ref{14})  match with the ones reported in \cite{las2017quantum}. The variation of $G$ with respect to various parameters is plotted in Fig.~\ref{fig:3}. To outperform the classical illumination, the condition that must be satisfied is $G > 1$. We focus our study on the case where the classical protocol yields the best result, specifically when \( \cos^2 \phi = 1 \). We call this case the `worst-case scenario' for QI.

We begin by examining Fig.~\ref{fig:3a}, which shows the variation of $G$ as a function of the signal strength $ N$ for different values of the target reflectivity $p$. In the weak-signal regime ($N \ll 1$), a significant advantage over the classical protocol is observed, with the gain reaching values as high as $G \simeq 2$, corresponding to an improvement of approximately $3\,\mathrm{dB}$, as quantified by $10 \log_{10}(G)$. This regime is particularly relevant for QI, where the goal is to detect a weakly reflecting target while minimizing the likelihood of revealing the presence of the probing signal.

\begin{figure}
    \centering
    \includegraphics[width=\linewidth]{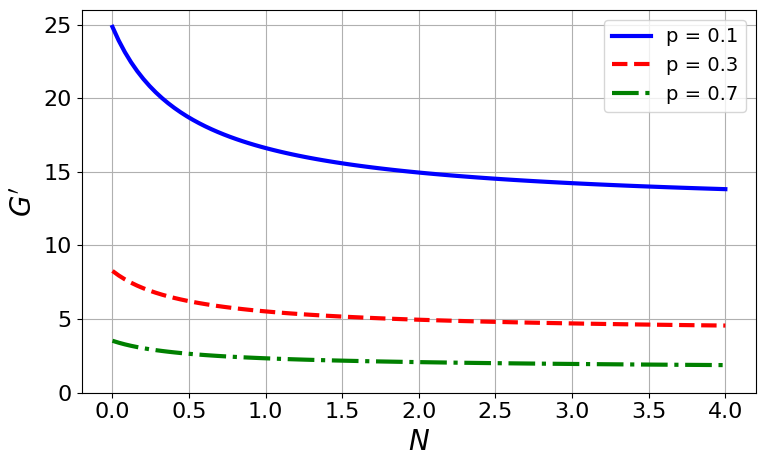}
    \caption{Phase-averaged gain $G^\prime$ as a function of the signal strength $N$ for different values of $p$. All the other parameters are the same as in Fig.~\ref{fig:3}.}
    \label{fig:4}
\end{figure}

\begin{figure*}

\centering
\subfloat[]{%
    \includegraphics[width=0.45\textwidth]{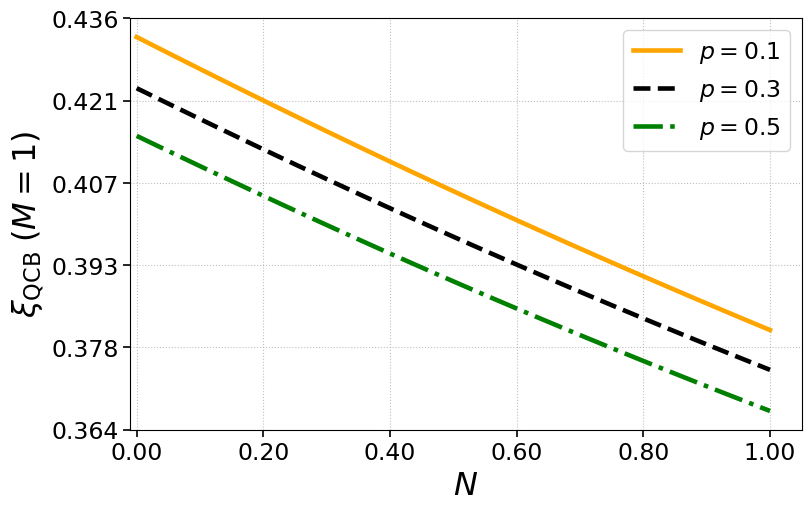}
    \label{fig:5a}
}
\hfill
\subfloat[]{%
    \includegraphics[width=0.45\textwidth]{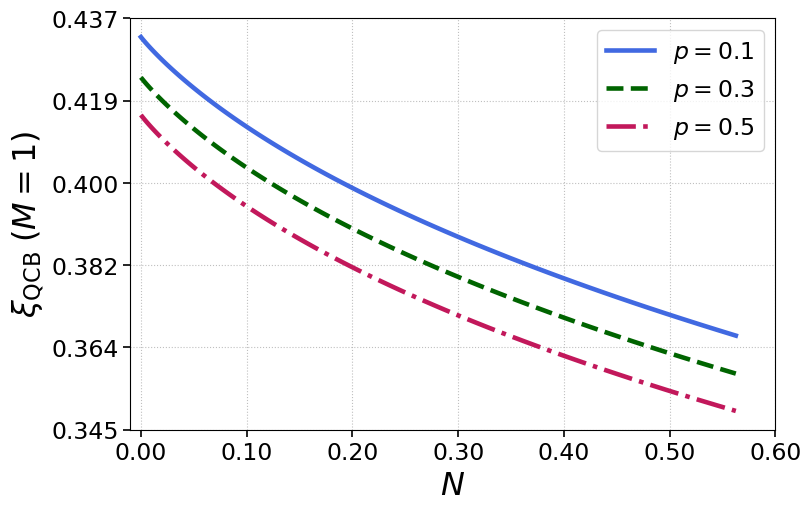}
    \label{fig:5b}
}

\caption{Variation of the QCB as a function of the average signal photon number $N$ for (a) the CS and (b) the TMSS. The other parameters are $\eta = 0.8$, $n_{th} = 3$, and $n_t = 4$.}
\label{fig:5}

\end{figure*}

Counterintuitively, \( G \) increases as \( p \) decreases, indicating that the quantum protocol is more effective at detecting lossy targets. As also discussed in \cite{gupta2021quantum}, this advantage arises because, together with the signal, the environmental thermal noise that was mixed with the signal is also partially lost upon interaction with a weakly reflecting target. Even in the opposite limit of strong signal strength ($N \gg 1$), the quantum protocol continues to outperform its classical counterpart for $p < 0.8$, with $G>1$, in contrast to the case of a perfectly reflecting target ($p=1$), where the quantum advantage disappears in this regime. 

Figs.~\ref{fig:3b} and~\ref{fig:3c} further demonstrate the robustness of the quantum protocol against thermal noise originating from both the environment and the target. The mean thermal photon numbers $n_{th}$ and $n_t$ are defined as $n = \mathrm{Tr}(\rho\, a^\dagger a)$, where $\rho$ denotes the thermal state. They follow the Bose--Einstein distribution, $n = 1/(\exp(\hbar\omega/k_B T) - 1)$, corresponding to the environmental and target temperatures, respectively, which may differ slightly. For a broad range of frequencies, spanning from the optical to the microwave regime, these quantities can take values ranging from nearly 0 to $\sim 110$. In Fig.~\ref{fig:3b}, the gain $G$ exhibits only a weak dependence on $n_{th}$, remaining nearly constant even for large thermal backgrounds. This highlights the resilience of quantum illumination in the presence of environmental noise, even in the microwave region. Further, as shown in Fig.~\ref{fig:3c},  $G$ increases with $n_t$ before saturating at higher values, indicating that thermal noise associated with the target can enhance the distinguishability between the target-present and target-absent hypotheses. Importantly, the gain remains greater than unity across the entire range of $n_{th}$ and $n_t$ considered, demonstrating the reliability of the quantum protocol over a wide range of operating frequencies relevant to radar and lidar applications.

The influence of the environmental reflectivity $\eta$ on the gain is shown in Fig.~\ref{fig:3d}. As $\eta$ increases, a slight but systematic decrease in $G$ is observed, reflecting the expected degradation of performance due to increased coupling with a bright thermal background. Nevertheless, the gain remains above unity over the relevant parameter range, confirming the persistence of the quantum advantage in the weak-signal regime. Fig.~\ref{fig:3e} illustrates the dependence of the gain on the target reflectivity $p$. The gain decreases monotonically with increasing $p$, consistent with the trends observed in Fig.~\ref{fig:3a}. This confirms that the proposed scheme optimally performs for weakly reflecting targets, where classical detection strategies are least effective. 
Finally, Fig.~\ref{fig:3f} presents the phase dependence of the gain as a function of the phase shift $\phi$ for a fixed signal strength $N=0.1$. The gain shows a periodic variation with $\phi$, yet remains consistently above unity across the full phase-shift interval of the returned signal. The $G$ is maximized at $\phi=\pm\pi/2$, where $\cos^2\phi=0$ and the denominator in Eq.~\ref{14} attains its minimum value, while it is slightly reduced at $\phi=0$ and $\phi=\pm\pi$, where $\cos^2\phi=1$.

In practical scenarios, the phase shift $\phi$ acquired by the signal may not be known or may fluctuate randomly due to environmental uncertainties. To account for this, we consider a phase-averaged performance by integrating the respective SNRs over all possible values of $\phi$ in the interval $[-\pi, \pi]$. This approach effectively models situations where no prior phase information is available at the detection stage. We define the averaged gain as
\begin{equation}
G' = \frac{\int_{-\pi}^{\pi} \mathrm{SNR}_Q(\phi)\, d\phi}{\int_{-\pi}^{\pi} \mathrm{SNR}_C(\phi)\, d\phi}.
\end{equation}
As shown in Fig.~\ref{fig:4}, the gain $G'$ remains significantly greater than unity over a broad range of $N$, indicating a clear quantum advantage even after phase averaging. This demonstrates the quantum protocol's reliability against both phase uncertainty and variations in target reflectivity. Furthermore, $G'$ is enhanced for smaller values of the target reflectivity $p$ and signal strength $N$, with the maximum value reaching up to $25$, corresponding to an improvement of approximately $14\,\mathrm{dB}$ in the absence of phase information. 

\section{COMPARATIVE STUDY FOR THE QCB}
\label{IV}

The evaluation of the QCB is carried out for a single copy of the probe, i.e., $M=1$, since the Chernoff bound is asymptotically tight and the bound approaches equality in the large $M$ limit. The numerical calculations required to evaluate Eq.~(\ref{9}) are done using QuTiP~\cite{lambert2026qutip}, following the methodology outlined in \cite{fan2018quantum}. The density matrices corresponding to the states in Eqs.~(\ref{10}) and (\ref{12}) were first constructed, while the thermal states \( \rho_h \) and \( \rho_t \) were generated using the built-in functions available in QuTiP. Throughout the calculation, the Hilbert space of all states was truncated to a dimension of 20 to ensure a feasible runtime, given the high computational complexity of the model. Due to this truncation constraint, the average photon number in the signal mode of the probe state was limited up to approximately 1 for CS and 0.56 for the TMSS. Furthermore, the thermal photon numbers \( n_{th} \) and \( n_t \) were restricted to values of about 3 and 4, respectively. It is important to emphasize that these values of $n_{th}$ and ${n_t}$ do not correspond to the microwave regime; consequently, the present QCB analysis is not carried out in the microwave domain. 

The variation of the QCB with the average signal photon number $N$ for both probe states is shown in Fig.~\ref{fig:5}. In both cases, the QCB decreases significantly with increasing $N$, indicating a corresponding reduction in the error probability. This behavior can be attributed to the additional thermal contribution from the target, which mixes with the signal and enhances the distinguishability between the returned signal and the background noise, thereby improving discrimination between the target-present and target-absent scenarios. For a given signal power, the TMSS is observed to attain a lower QCB more rapidly than the CS as $N$ increases, leading to a lower error probability. This demonstrates that the TMSS serves as a reliable QI source in low-signal-strength regimes.

This striking decrease in the value of the QCB can be explained in the following way. Our sequential model introduces a physically distinct and crucial feature: the target is associated with its own independent thermal mode, characterized by an average photon number $n_t $, which is separate from the environmental thermal mode with $n_{th}$. In the absence of the target, the measurement device effectively receives only the environmental noise leaking through the interaction, dominated by the mode $\sqrt{1-\eta}\, a_h$. In contrast, when the target is present, the returned signal contains contributions from both the environmental and target thermal modes, mixed through two successive and independent BS. Since $n_t \neq n_{th}$, the intrinsic thermal emission of the target provides an additional distinguishing signature between the two hypotheses, thereby enhancing the separability of the states $\rho_0$ and $\rho_1$. This enhanced distinguishability is reflected in the behavior of the QCB, which decreases significantly from $\xi_{\mathrm{QCB}} \approx 0.43$ to $\approx 0.35$, for the TMSS, over the range $N \in [0,\, 0.56]$. In contrast, the model of ~\cite{fan2018quantum}, which employs a single BS and assumes the target to be in thermal equilibrium with the environment, exhibits very low variation in the QCB over the range $N \in [0,\, 2]$, and the corresponding output states $\rho_0$ and $\rho_1$ remain nearly indistinguishable. It has been established in~\cite{fan2018quantum} that, for a fixed mean photon number in the signal mode, the TMSS achieves the lowest QCB when compared with non-Gaussian states obtained via photon addition and subtraction. In this context, the lower QCB obtained in our model for the TMSS, relative to previously reported values, further reinforces its optimality as a probe state under constraints on the signal strength.

\section{Conclusion}
\label{V}

In this work, the QI protocol has been modeled by incorporating the sequential interaction of the signal with a noisy environment, followed by its interaction with the target. This model is more realistic than the ones used in literature in the sense that it allows us to explicitly incorporate the effect of a target that is both lossy and not in thermal equilibrium with its surroundings. The target is modeled as a low-reflectivity beam splitter with its own thermal contribution, independent of the environmental thermal noise. Within this framework, a comparative study of CI, employing CS, and QI, employing the TMSS, has been carried out using two performance metrics: the SNR and the QCB. 

The analysis shows that, despite the presence of additional thermal noise from the target and attenuation of the signal due to its low reflectivity, the TMSS continues to outperform the CS, not only in the low photon-number regime but also for moderate signal strengths (e.g., $N \approx 4$). Moreover, the performance gain is more pronounced for low-reflectivity targets than for perfectly reflecting ones, reaching values as high as $3\,\mathrm{dB}$. This indicates that QI with TMSS is particularly advantageous for detecting lossy targets. In addition, to account for the scenarios in which the phase shift acquired by the signal may be unknown or fluctuating, we have analyzed the phase-averaged performance by integrating the SNR over all possible values of $\phi$. The resulting averaged gain remains significantly greater than unity over a broad range of signal strength of the probe state, reaching values up to $14\,\mathrm{dB}$ for highly lossy targets. This demonstrates that the quantum advantage persists even in the absence of phase information.

Furthermore, within the proposed model, the QCB is found to decrease more rapidly with increasing signal strength, even for a single copy of the probe, indicating enhanced distinguishability between the target-present and target-absent hypotheses. The TMSS achieves lower QCB values at a given average signal photon number than the corresponding single-bath setting. This is primarily because the target’s independent thermal mode provides an additional channel for distinguishability. Overall, the present model provides a more comprehensive description of the target detection process and offers a pathway for incorporating additional sources of attenuation, as well as for exploring the role of other non-classical states for the proposed model, thereby providing a promising direction for future studies in realistic quantum sensing.

\section*{Acknowledgements}
The authors express their sincere gratitude to Dr. M. Sanz and Mr. Jayesh Shukla for their valuable insights and constructive discussions.

\bibliographystyle{ieeetr}
\bibliography{main}

\end{document}